\documentstyle[12pt]{book} %%%%% IBM Research Report RC XXXXX
\begin{document}

\sloppy
\raggedbottom

\begin{titlepage}

\hfill\vspace{1in}\\
{\Huge\bf\hspace{-\parindent}The Limits of \vspace{3mm}\\
Mathematics (in C) \vspace{1.5in} \\
}
{\Large\bf G J Chaitin \vspace{1mm} \\
IBM, P O Box 704 \\
Yorktown Heights, NY 10598 \vspace{1mm} \\
{\it chaitin@watson.ibm.com} \vspace{1.5in} \\
April 27, 1994
}

\end{titlepage}

\begin{titlepage}

\hfill\\

\end{titlepage}

\markboth
{The Limits of Mathematics (in C)}{}

\chapter*{Preface}

In a remarkable development, I have constructed a new definition for a
self-delimiting universal Turing machine (UTM) that is easy to program
and runs very quickly.  This provides a new foundation for algorithmic
information theory (AIT), which is the theory of the size in bits of
programs for self-delimiting UTM's.  Previously, AIT had an abstract
mathematical quality.  Now it is possible to write down executable
programs that embody the constructions in the proofs of theorems.  So
AIT goes from dealing with remote idealized mythical objects to being
a theory about practical down-to-earth gadgets that one can actually
play with and use.

This new self-delimiting UTM is implemented via software written in
C that is based on the related and now largely
obsolete software that I presented in my report ``Exhibiting
randomness in arithmetic using Mathematica and C,'' IBM Research
Report RC-18946, 94 pp., June 1993.  In its turn this report was a
reworking of the software for my book {\it Algorithmic Information
Theory,} Cambridge University Press, 1987.

Using this new software, as well as the latest theoretical ideas, it
is now possible to give a self-contained ``hands on'' mini-course
presenting very concretely my latest proofs of my two fundamental
information-theoretic incompleteness theorems.  The first of these
theorems states that an $N$-bit formal axiomatic system cannot enable
one to exhibit any specific object with program-size complexity
greater than $N+c$.  The second of these theorems states that an
$N$-bit formal axiomatic system cannot enable one to determine more
than $N+c'$ scattered bits of the halting probability $\Omega$.

Most people believe that anything that is true is true for a reason.
These theorems show that some things are true for no reason at all,
i.e., accidentally, or at random.

The latest and I believe the deepest proofs of these two theorems were
originally presented in my paper ``Information-theoretic
incompleteness,'' {\it Applied Mathematics and Computation\/} 52
(1992), pp.\ 83--101.  This paper is reprinted in my book {\it
Information-Theoretic Incompleteness,} World Scientific, 1992.

As is shown in this course, the algorithms considered in the
proofs of these two theorems are now easy to program and run, and by
looking at the size in bits of these programs one can actually, for
the first time, determine exact values for the constants $c$ and $c'$.

It is my intention to use this approach and software in an intensive
short course on the limits of mathematics that I will give at the
University of Maine in Orono
in the summer of 1994.  I also intend to write up this
course as a new book.  This research report, however, is intended to
make these important new ideas and software widely available in
preliminary form as soon as possible.

I wish to thank Prof.\ George Markowsky of the University of Maine at
Orono for his stimulating invitation to give a summer workshop.  I
also thank Prof.\ John Casti of the Santa Fe Institute, Prof.\ Carlton
Caves of the University of New Mexico, and Prof.\ Cristian Calude of
the University of Auckland for stimulating discussions.

I am grateful to IBM for its enthusiastic and unwavering support of my
research for a quarter of a century, and to my current management
chain at the IBM Research Division, Marty Hopkins, Eric Kronstadt, and
Jeff Jaffe.  Finally I thank the RISC project group, of which I am a
member, for designing the marvelous IBM RISC System/6000 workstations
that I have used for all these calculations, and for providing me with
the latest models of this spectacular computing equipment.

All enquires, comments and suggestions regarding this software should
be sent via e-mail to {\tt chaitin} at {\tt watson.ibm.com}.

\tableofcontents

\newcommand
{\chap}[1]{\chapter*{#1}\markboth{The Limits of Mathematics (in C)}{#1}
\addcontentsline{toc}{chapter}{#1}}
\newcommand
{\Size}{\small}   % \small \footnotesize \scriptsize

\part{Explanation}

\chap{The New Idea}

First of all, note that we use here the toy LISP from my monograph
{\it Algorithmic Information Theory,} Cambridge University Press,
1987.  (Reprinted with minor revisions thrice, lastly in 1992.)

Here is a quick summary of this toy LISP.  Each LISP primitive
function and variable is as before a single character, but they have
been changed from the IBM APL2 character set to the standard ASCII
character set.  These primitive functions, all of which have a fixed
number of arguments, are now {\tt '} for QUOTE (1 argument), {\tt .}
for ATOM (1 argument), {\tt =} for EQ (2 arguments), {\tt +} for CAR
(1 argument), {\tt -} for CDR (1 argument), {\tt *} for CONS (2
arguments), {\tt \&} for LAMBDA and DEFINE (2 arguments), {\tt :} for
LET-BE-IN (3 arguments), {\tt /} for IF-THEN-ELSE (3 arguments), {\tt
,} for OUTPUT (1 argument), {\tt !} for EVAL (1 argument), and {\tt ?}
for depth-limited EVAL (had 2 arguments, now has 3).  The
meta-notation {\tt "} indicates that an S-expression with explicit
parentheses follows, not what is usually the case in my toy LISP, an
M-expression, in which the parentheses for each primitive function are
implicit.  Finally the empty list NIL is written {\tt ()}, and TRUE
and FALSE are {\tt 1} and {\tt 0}.

The new idea that goes beyond what is presented in my Cambridge
University Press monograph is this.  We define our standard
self-delimiting universal Turing machine as follows.  Its program is
in binary, and appears on a tape in the following form.  First comes a
toy LISP expression, written in ASCII with 7 bits per character.  Note
that this expression is self-delimiting because parentheses must
balance.  The TM reads in this LISP expression, and then evaluates it.
As it does this, two new primitive functions {\tt @} and {\tt \%} with
no arguments may be used to read more from the TM tape.  Both of these
functions explode if the tape is exhausted, killing the computation.
{\tt @} reads a single bit from the tape, and {\tt \%} reads in an
entire LISP expression, in 7-bit character chunks.

This is the only way that information on the TM tape may be accessed,
which forces it to be used in a self-delimiting fashion.  This is
because no algorithm can search for the end of the tape and then use
the length of the tape as data in the computation.  If an algorithm
attempts to read a bit that is not on the tape, the algorithm aborts.

How is information placed on the TM tape in the first place?  Well, in
the starting environment, the tape is empty and any attempt to read it
will give an error message.  To place information on the tape, a new
argument has been added to the primitive function {\tt ?} for
depth-limited evaluation.

Consider the three arguments $\alpha$, $\beta$ and $\gamma$ of {\tt
?}.  The meaning of the first argument, the depth limit $\alpha$, has
been changed slightly.  If $\alpha$ is a non-null atom, then there is
no depth limit.  If $\alpha$ is the empty list, this means zero depth
limit (no function calls or re-evaluations).  And an $N$-element list
$\alpha$ means depth limit $N$.  The second argument $\beta$ of {\tt
?} is as before the expression to be evaluated as long as the depth
limit $\alpha$ is not exceeded.  The new third argument $\gamma$ of
{\tt ?} is a list of bits to be used as the TM tape.

The value $\nu$ returned by the primitive function {\tt ?} is also
changed.  $\nu$ is a list.  The first element of $\nu$ is {\tt !} if
the evaluation of $\beta$ aborted because an attempt was made to read
a non-existent bit from the TM tape.  The first element of $\nu$ is
{\tt ?} if evaluation of $\beta$ aborted because the depth limit
$\alpha$ was exceeded.  {\tt !} and {\tt ?} are the only possible
error flags, because my toy LISP is designed with maximally permissive
semantics.  If the computation $\beta$ terminated normally instead of
aborting, the first element of $\nu$ will be a list with only one
element, which is the result produced by the computation $\beta$.
That's the first element of the list $\nu$ produced by the {\tt ?}
primitive function.

The rest of the value $\nu$ is a stack of all the arguments to the
primitive function {\tt ,} that were encountered during the evaluation
of $\beta$.  More precisely, if {\tt ,} was called $N$ times during
the evaluation of $\beta$, then $\nu$ will be a list of $N+1$
elements.  The $N$ arguments of {\tt ,} appear in $\nu$ in inverse
chronological order.  Thus {\tt ?} can not only be used to determine
if a computation $\beta$ reads too much tape or goes on too long
(i.e., to greater depth than $\alpha$), but {\tt ?} can also be used
to capture all the output that $\beta$ displayed as it went along,
whether the computation $\beta$ aborted or not.

In summary, all that one has to do to simulate a self-delimiting
universal Turing machine $U(p)$ running on the binary program $p$ is
to write
\begin{verbatim}
                         ?0'!%p
\end{verbatim}
This is an M-expression with parentheses omitted from primitive
functions, because all primitive functions have a fixed number of
arguments.  With all parentheses supplied, it becomes the S-expression
\begin{verbatim}
                     (?0('(!(%)))p)
\end{verbatim}
This says that one is to read a complete LISP S-expression from the TM
tape $p$ and then evaluate it without any time limit and using
whatever is left on the tape $p$.

Two more primitive functions have also been added, the 2-argument
function \verb|^| for APPEND, i.e., list concatenation, and the
1-argument function {\tt \#} for converting an expression into the
list of the bits in its ASCII character string representation.  These
are used for constructing the bit strings that are then put on the TM
tape using {\tt ?}'s third argument $\gamma$.  Note that the functions
\verb|^|, {\tt \#} and {\tt \%} could be programmed rather than
included as built-in primitive functions, but it is extremely
convenient and much much faster to provide them built in.

Finally a new 1-argument identity function \verb|~| with the
side-effect of outputting its argument is provided for debugging.
Output produced by \verb|~| is invisible to the ``official'' {\tt ?}
and {\tt ,} output mechanism.  \verb|~| is needed because {\tt
?}$\alpha\beta\gamma$ suppresses all output $\theta$ produced within
its depth-controlled evaluation of $\beta$.  Instead {\tt ?} stacks
all output $\theta$ from within $\beta$ for inclusion in the final
value $\nu$ that {\tt ?} returns, namely $\nu = $ (atomic error flag
or (value of $\beta$) followed by the output $\theta$).

\chap{Course Outline}

The course begins by explaining with examples my toy LISP.  Possibly
the theory of LISP program-size complexity is developed a little,
following my recent papers ``LISP program-size complexity,'' {\it
Applied Mathematics and Computation\/} 49 (1992), pp.\ 79--93, ``LISP
program-size complexity II,'' {\it Applied Mathematics and
Computation\/} 52 (1992), pp.\ 103--126, ``LISP program-size
complexity III,'' {\it Applied Mathematics and Computation\/} 52
(1992), pp.\ 127--139, ``LISP program-size complexity IV,'' {\it
Applied Mathematics and Computation\/} 52 (1992), pp.\ 141--147.
These papers are reprinted in my book {\it Information-Theoretic
Incompleteness,} World Scientific, 1992.

LISP program-size complexity is extremely simple and concrete.  In
particular, it is easy to show that it is impossible to prove that a
self-contained LISP expression is elegant, i.e., that no smaller
expression has the same value.  To prove that an $N$-character LISP
expression is elegant requires a formal axiomatic system that itself
has LISP complexity $N$.  Also, LISP program-size complexity is
subadditive, because expressions are self-delimiting and can be
concatenated, and also because we are dealing with pure LISP and no
side-effects get in the way.  Moreover, the probability $\Omega_{\rm
LISP}$ that a LISP expression ``halts'' or has a value is
well-defined, also because programs are self-delimiting.  Finally, it
is easy to see that the LISP program-size complexity of the first $N$
bits of the LISP halting probability $\Omega_{\rm LISP}$ grows
linearly with $N$.  Therefore to be able to determine the first $N$
bits of $\Omega_{\rm LISP}$ requires a formal axiomatic system whose
LISP complexity also grows linearly with $N$.

It should then be pointed out that LISP programs have severe
information-theoretic limitations because they do not encode
information very efficiently in 7-bit ASCII characters subject to LISP
syntax constraints.  Arbitrary binary programs are denser and much
better, but they should at least be kept self-delimiting.

So next we define our standard self-delimiting universal Turing
machine $U(p)$ using
\begin{verbatim}
                         ?0'!%p
\end{verbatim}
as explained in the previous chapter.

Next we show that
\[
   H(x,y) \le H(x) + H(y) + c.
\]
Here $H(\cdots)$ denotes the size in bits of the smallest program that
makes our standard universal Turing machine compute $\cdots$.  Thus
this inequality states that the information needed to compute the pair
$(x,y)$ is bounded by a constant $c$ plus the sum of the information
needed to compute $x$ and the information needed to compute $y$.
Consider
\begin{verbatim}
                        *!%*!%()
\end{verbatim}
This is an M-expression with parentheses omitted from primitive
functions.  With all parentheses supplied, it becomes the S-expression
\begin{verbatim}
                  (*(!(%))(*(!(%))()))
\end{verbatim}
In fact, $c$ is just 7 times the size in characters of this LISP
S-expression, which is exactly 20 characters.  Thus $c = 7 \times 20 =
140$ bits!  See {\tt univ.lisp}.  Note that in standard LISP this
would be something like
\begin{verbatim}
              (CONS (EVAL (READ-EXPRESSION))
              (CONS (EVAL (READ-EXPRESSION))
                    NIL))
\end{verbatim}
which is much more than 20 characters long.

Looking at binary strings $x$ with size $|x|$ in bits, we next show
that
\[
   H(x) \le 2|x| + c
\]
and
\[
   H(x) \le |x| + H(|x|) + c'.
\]
As before, the programs for doing this are exhibited and run.
Also $c$ and $c'$ are determined.
See {\tt univ.lisp}.

Next we show how to calculate the halting probability $\Omega$ of our
standard self-delimiting universal Turing machine in the limit from
below.  The LISP program for doing this, {\tt omega.lisp}, is now
remarkably clear and fast, and much better than the one given in my
Cambridge University Press monograph.  (See also {\tt omega2.lisp} and
{\tt omega3.lisp}.)  Using this, we show that if $\Omega_N$ denotes
the first $N$ bits of the fractional part of the base-two real number
$\Omega$, then
\[
   H(\Omega_N) \ge N - c.
\]
Again this is done with a program that can actually be run and whose
size gives us a value for $c$.  See {\tt omega4.lisp}.

Next we turn to the self-delimiting program-size complexity $H(X)$ for
infinite r.e.\ sets $X$, which is not considered at all in my
Cambridge University Press monograph.  This is now defined to be the
size in bits of the smallest LISP expression $\xi$ that executes
forever without halting and outputs the members of the r.e.\ set $X$
using the LISP primitive {\tt ,}.  {\tt ,} is an identity function
with the side-effect of outputting the value of its argument.  Note
that this LISP expression $\xi$ is allowed to read additional bits or
expressions from the TM tape using the primitive functions {\tt @} and
{\tt \%} if $\xi$ so desires.  But of course $\xi$ is charged for
this; this adds to $\xi$'s program size.

It was in order to deal with such unending expressions $\xi$ that the
LISP primitive function for time-limited evaluation {\tt ?} now
captures all output from {\tt ,} within its second argument $\beta$.

To illustrate these new concepts, we show that
\[
   H(X \cap Y) \le H(X) + H(Y) + c
\]
and
\[
   H(X \cup Y) \le H(X) + H(Y) + c'
\]
for infinite r.e.\ sets $X$ and $Y$.  As before, we run examples and
determine $c$ and $c'$.  See {\tt sets0.lisp} through {\tt
sets4.lisp}.

Now consider a formal axiomatic system $A$ of complexity $N$,
i.e., with a set of theorems $T_A$ that considered as an r.e. set as
above has self-delimiting program-size complexity $H(T_A)$.  We show
that $A$ has the following limitations.  First of all, we show
directly that $A$ cannot enable us to exhibit a specific S-expression
$s$ with self-delimiting complexity $H(s)$ greater than $N+c$.  See
{\tt godel.lisp} and {\tt godel2.lisp}.

Secondly, using the lower bound of $N-c$ on $H(\Omega_N)$ established
in {\tt omega4.lisp}, we show that $A$ cannot enable us to determine
not only the first $N+c'$ bits of $\Omega$, but any $N+c'$ bits of
$\Omega$, even if they are scattered and we leave gaps.  (See {\tt
godel3.lisp}.)  In my Cambridge University Press monograph, this took
a hundred pages to show, and involved the systematic development of a
general theory using measure-theoretic arguments.  Following
``Information-theoretic incompleteness,'' {\it Applied Mathematics and
Computation\/} 52 (1992), pp.\ 83--101, the proof is now a
straight-forward Berry-paradox-like program-size argument.  Moreover
we are using a deeper definition of $H(A) \equiv H(T_A)$ via infinite
self-delimiting computations.

And last but not least, the philosophical implications of all this
should be discussed, especially the extent to which it tends to
justify experimental mathematics.  This would be along the lines of
the discussion in my talk transcript ``Randomness in arithmetic and
the decline and fall of reductionism in pure mathematics,'' {\it
Bulletin of the European Association for Theoretical Computer
Science,} No.\ 50 (June 1993), pp.\ 314--328.

This concludes our ``hand-on'' mini-course on the
information-theoretic limits of mathematics.

\chap{Software User Guide}

All the software for this course is written in a toy version of {\sl LISP}.
{\tt *.lisp} files are toy {\sl LISP} code.
One {\sl C} program, {\tt lisp.c,}
is also provided; it is the {\sl LISP} interpreter.
In these instructions we assume that this software is being run in a
{\sl UNIX} environment.

To run the {\sl LISP} interpreter, first compile it.
This is done using the command {\tt cc -O -olisp lisp.c}.
This interpreter is about 80 megabytes big.  If this is too large,
reduce the \verb|#define SIZE| before compiling it.

To run the {\sl LISP} interpreter interactively, that is, with
input and output on the screen, enter
\[
\mbox{\tt lisp}
\]
As
each M-expression is read in, it is written out, then converted to an
S-expression that is written out and evaluated.\footnote{The
conversion from M-to S-expression mostly consists of making all
implicit parentheses explicit.}

To run a {\sl LISP} program {\tt X.lisp} with output on the screen, enter
\[
\mbox{\tt lisp <X.lisp}
\]
To run a {\sl LISP} program {\tt X.lisp} with output in a file
rather than on the screen, enter
\[
\mbox{\tt lisp <X.lisp >X.run}
\]

\chap{Bibliography}

\begin{itemize}
\item[{[1]}] B. W. Kernighan and D. M. Ritchie, {\it The C Programming
Language,} second edition, Prentice Hall, 1988.
\item[{[2]}] G. J. Chaitin, {\it Algorithmic Information Theory,}
revised third printing, Cambridge University Press, 1990.
\item[{[3]}] G. J. Chaitin, {\it Information, Randomness \&
Incompleteness,} second edition, World Scientific, 1990.
\item[{[4]}] G. J. Chaitin, ``LISP program-size complexity,'' {\it
Applied Mathematics and Computation\/} 49 (1992), pp.\ 79--93.
\item[{[5]}] G. J. Chaitin, ``Information-theoretic incompleteness,''
{\it Applied Mathematics and Computation\/} 52 (1992), pp.\ 83--101.
\item[{[6]}] G. J. Chaitin, ``LISP program-size complexity II,'' {\it
Applied Mathematics and Computation\/} 52 (1992), pp.\ 103--126.
\item[{[7]}] G. J. Chaitin, ``LISP program-size complexity III,'' {\it
Applied Mathematics and Computation\/} 52 (1992), pp.\ 127--139.
\item[{[8]}] G. J. Chaitin, ``LISP program-size complexity IV,'' {\it
Applied Mathematics and Computation\/} 52 (1992), pp.\ 141--147.
\item[{[9]}] G. J. Chaitin, {\it Information-Theoretic
Incompleteness,} World Scientific, 1992.
\item[{[10]}] G. J. Chaitin, ``Randomness in arithmetic and the
decline and fall of reductionism in pure mathematics,'' {\it Bulletin
of the European Association for Theoretical Computer Science,} No.\ 50
(June 1993), pp.\ 314--328.
Reprinted in
{\it International Journal of Bifurcation and Chaos},
Vol.\ 4, No.\ 1 (February 1994), in press.
\item[{[11]}] G. J. Chaitin, ``Exhibiting randomness in arithmetic
using Mathematica and C,'' IBM Research Report RC-18946, June 1993.
\item[{[12]}] G. J. Chaitin, ``On the number of $n$-bit strings with
maximum complexity,'' {\it Applied Mathematics and Computation\/} 59
(1993), pp.\ 97--100.
\item[{[13]}] G. J. Chaitin, ``Randomness in arithmetic and the limits
of mathematical reasoning,'' in J. Tr\^an Thanh V\^an, {\it
Proceedings of the Vth Rencontres de Blois}, Editions Fronti\`eres, in
press.
\item[{[14]}] C. Calude, {\it Information and Randomness,}
Springer-Verlag, in press.
\item[{[15]}] K. Svozil, {\it Randomness \& Undecidability in
Physics,} World Scientific, 1993.
\end{itemize}

\part{The Course}

{
% *.lisp
}\chap{univ.lisp}{\Size\begin{verbatim}
[[[
 First steps with my new construction for
 a self-delimiting universal Turing machine.
 We show that
    H(x,y) <= H(x) + H(y) + c
 and determine c.
 Consider a bit string x of length |x|.
 We also show that
    H(x) <= 2|x| + c
 and that
    H(x) <= |x| + H(|x|) + c
 and determine both these c's.
]]]

[first demo the new lisp primitive functions]
^'(1234567890)'(abcdefghi)
@
?0 '@ '()
?0 '@ '(1)
?0 '@ '(0)
?0 '@ '(x)
?0 '**@()**@()() '(10)
?0 '**,@()**,@()() '(10)
?0 '**,@()**,@()**,@()() '(10)
#'a
#'(abcdef)
#'(12(34)56)
?0 '% '(110 0001)
?0 '% '(110 0010)
?0 '% '(110 0011)
?0 '% '(110 0100)
?0 '% '(110 0101)
?0 '% '(110 0110)
?0 '% '(110 0111)
?0 '% #'a
?0 '% '(010 100)
?0 '% '(010 1001)
?0 '% '(010 1000 011 0001 011 0010 010 1001)
?0 '% '(010 1000 011 0001 011 0010         )
?0 '% '(010 1000 011 0001 011 001          )
?0 '% '(010 100                            )
?0 '% #'(abcdef)
?0 '% #'(12(34)56)
?0 '*%*%() ,^ #'a #'b
?0
':(f) :x@ :y@ /=xy *x(f) () (f)
'(0011001101)
#':(f) :x@ :y@ /=xy *x(f) () (f)
[ Here is the self-delimiting universal Turing machine! ]
[ (with slightly funny handling of out-of-tape condition) ]
& (Up) ++?0'!%p
[Show that H(x) <= 2|x| + c]
(U
 ^ ,#,':(f) :x@ :y@ /=xy *x(f) () (f)
   '(0011001101)
)
(U
 ^ ,#,':(f) :x@ :y@ /=xy *x(f) () (f)
   '(0011001100)
)
(U
 ^ ,#,'*!%*!%() [The length of this bit string is the]
                [constant c in H(x,y) <= H(x)+H(y)+c.]
 ^ #':(f) :x@ :y@ /=xy *x(f) () (f)
 ^ '(0011001101)
 ^ #':(f) :x@ :y@ /=xy *x(f) () (f)
   '(1100110001)
)
[Size of list in reverse decimal!]
& (Se) /.e() (I[,](S-e))
[Increment reverse decimal]
& (In) /.n'(1) :d+n :r-n
     /=d0*1r
     /=d1*2r
     /=d2*3r
     /=d3*4r
     /=d4*5r
     /=d5*6r
     /=d6*7r
     /=d7*8r
     /=d8*9r
    [/=d9]*0(Ir)
[Reverse list]
& (Re) /.e() ^(R-e)*+e()
[Convert to binary and display size in decimal]
& (Me) :e [,]#[,]e :f ,(R[,](Se)) e
(M'a)
(M'())
& (Dk) /=0+k *1(D-k) /.-k () *0-k [D = decrement]
,(D,(D,(D,(D,'(001)))))
(U
 ^ ,#,'   [Show that H(x) <= |x| + H(|x|) + c]
   : (Re) /.e() ^(R-e)*+e()          [R = reverse  ]
   : (Dk) /=0+k *1(D-k) /.-k () *0-k [D = decrement]
   : (Lk) /.k () *@(L(Dk))           [L = loop     ]
   (L(R!%))
 ^ #''(1000)
   '(0000 0001)
)
\end{verbatim}
}\chap{omega.lisp}{\Size\begin{verbatim}

[[[[ Omega in the limit from below! ]]]]

[Look at the behavior of typical 7-bit programs]
?0'!,%'(010 1000) [lpar]
?0'!,%'(010 1001) [rpar]
?0'!,%'(011 0001) [1]
[All strings of length k / with same length as k ]
& (Xk) /.k'(()) (Z(X-k))
[Append 0 and 1 to each element of list]
& (Zx) /.x() *^+x'(0) *^+x'(1) (Z-x)
(Z'((a)(b)))
(X'())
(X'(1))
(X'(11))
(X'(111))
[Size of list in reverse binary]
& (Se) /.e() (I(S-e))
[Increment reverse binary]
& (Ix) /.x'(1) /=+x0 *1-x *0(I-x)
(S'())
(S'(a))
(S'(ab))
(S'(abc))
(S'(abcd))
[Pad x to length of k on right and reverse]
& (Rxk) /.x /.k() *0(Rx-k) ^(R-x-k)*+x()
(R'(1)'(11))
(R'(01)'(1111))
(R'(0001)'(1111 1111))
[Set of programs in x that halt within time k]
& (Hxk) /.x() /,=0.+?k'!%,+x *+x(H-xk) (H-xk)
(H '((111)(111 1111)(000)(000 0000)) '0)
(H , *#'a *#'('(xy)) *#':(X)(X)(X) () '(111))
[For LISP omega must separate read & exec.  ]
& (Gxk) /.x()  [version of H for lisp omega ]
 : e  ?0'%,+x  [read expression from prog +x]
 [If read finished, evaluate exp for time k ]
 [with empty tape, so @ and % will fail!    ]
 : v  /.+e e ?k++e()[run for time k, no tape]
 /,=0.+v  *+x(G-xk) [program +x halted      ]
          (G-xk)    [program +x didn't halt ]
(G '((111)(111 1111)(000)(000 0000)) '0)
(G , *#'a *#'('(xy)) *#':(X)(X)(X) () '(111))
(H , *^#'@'(1) *^#'%#'(ab) () '(111))
(G , *^#'@'(1) *^#'%#'(ab) () '(111))
[Omega sub k!]
& (Wk) *0*".(R,(S,(H,(Xk)k))k)
(W'())
(W'(1))
(W'(11))
(W'(111))
[[[[
(W'(111 1))
(W'(111 11))
(W'(111 111))
]]]]
(W'(111 1111))
[[[[
(W'(111 1111 1))
]]]]
\end{verbatim}
}\chap{omega2.lisp}{\Size\begin{verbatim}

[[[[ Omega in the limit from below! ]]]]

[All strings of length k / with same length as k ]
& (Xk) /.k'(()) (Z(X-k))
[Append 0 and 1 to each element of list]
& (Zx) /.x() *^+x'(0) *^+x'(1) (Z-x)
[Size of list in reverse binary]
& (Se) /.e() (I(S-e))
[Increment reverse binary]
& (Ix) /.x'(1) /=+x0 *1-x *0(I-x)
[Pad x to length of k on right and reverse]
& (Rxk) /.x /.k() *0(Rx-k) ^(R-x-k)*+x()
[Set of programs in x that halt within time k]
& (Hxk) /.x() /=0.+?k'!%+x *+x(H-xk) (H-xk)
[Omega sub k!]
& (Wk) *0*".(R,(S,(H,(Xk)k))k)
(W'())
(W'(1))
(W'(11))
(W'(111))
(W'(111 1))
(W'(111 11))
(W'(111 111))
(W'(111 1111))
(W'(111 1111 1))
\end{verbatim}
}\chap{omega3.lisp}{\Size\begin{verbatim}

[[[[ Omega in the limit from below! ]]]]

[All strings of length k / with same length as k ]
& (Xk) /.k'(()) (Z(X-k))
[Append 0 and 1 to each element of list]
& (Zx) /.x() *^+x'(0) *^+x'(1) (Z-x)
[Size of list in reverse binary]
& (Se) /.e() (I(S-e))
[Increment reverse binary]
& (Ix) /.x'(1) /=+x0 *1-x *0(I-x)
[Pad x to length of k on right and reverse]
& (Rxk) /.x /.k() *0(Rx-k) ^(R-x-k)*+x()
[Set of programs in x that halt within time k]
& (Hxk) /.x() /=0.+?k'!%+x *+x(H-xk) (H-xk)
[Omega sub k!]
& (Wk) *0*".(R(S(H(Xk)k))k)
(W'())
(W'(1))
(W'(11))
(W'(111))
(W'(111 1))
(W'(111 11))
(W'(111 111))
(W'(111 1111))
(W'(111 1111 1))
(W'(111 1111 11))
(W'(111 1111 111))
(W'(111 1111 111 1))
(W'(111 1111 111 11))
(W'(111 1111 111 111))
(W'(111 1111 111 1111))
(W'(111 1111 111 1111 1))
(W'(111 1111 111 1111 11))
[The following exhaust storage:]
[[[[
(W'(111 1111 111 1111 111))
(W'(111 1111 111 1111 111 1))
(W'(111 1111 111 1111 111 11))
(W'(111 1111 111 1111 111 111))
(W'(111 1111 111 1111 111 1111))
]]]]
\end{verbatim}
}\chap{omega4.lisp}{\Size\begin{verbatim}

[[[
 Show that H(Omega sub n) > n - c and determine c.
 Omega sub n is the first n bits of Omega.
]]]

[First test new stuff]

[Compare two fractional binary numbers, i.e., is 0.x < 0.y ?]
& (<xy) /.x /.y   0
                  (<'(0)y)
            /.y   0
                  /+x /+y (<-x-y)
                          0
                      /+y 1
                          (<-x-y)
(<'(000)'(000))
(<'(000)'(001))
(<'(001)'(000))
(<'(001)'(001))
(<'(110)'(110))
(<'(110)'(111))
(<'(111)'(110))
(<'(111)'(111))
(<'()'(000))
(<'()'(001))
(<'(000)'())
(<'(001)'())

[Now run it all!]

++?0'!%

^,#,'

[All strings of length k / with same length as k ]
: (Xk) /.k'(()) (Z(X-k))
[Append 0 and 1 to each element of list]
: (Zx) /.x() *^+x'(0) *^+x'(1) (Z-x)
[Size of list in reverse binary]
: (Se) /.e() (I(S-e))
[Increment reverse binary]
: (Ix) /.x'(1) /=+x0 *1-x *0(I-x)
[Pad x to length of k on right and reverse]
: (Rxk) /.x /.k() *0(Rx-k) ^(R-x-k)*+x()
[Set of programs in x that halt within time k]
: (Hxk) /.x() /~=0.+?k'!%~+x *+x(H-xk) (H-xk)
[Omega sub k without 0. at beginning
 (i.e. only fractional part).]
: (Wk) (R(S(H(Xk)k))k)

[Compare two fractional binary numbers, i.e., is 0.x < 0.y ?]
: (<xy) /.x /.y   0
                  (<'(0)y)
            /.y   0
                  /+x /+y (<-x-y)
                          0
                      /+y 1
                          (<-x-y)
: w !%            [Read and execute from remainder of tape
                   a program to compute an n-bit
                   initial piece of Omega.]
: (Lk)            [Main Loop]
  : x    (Wk)     [Compute the kth lower bound on Omega]
  /(<xw) (L*1k)   [Are the first n bits OK?  If not, bump k.]
         (B(Xk))  [Form the union of all output of k-bit
                   programs within time k, output it,
                   and halt.
                   This is bigger than anything of complexity
                   less than or equal to n!]
[This total output will be bigger than each individual output,
 and therefore must come from a program with more than n bits.
]
[Union of all output of programs in list p within time k.]
: (Bp) /.p() * ~?k'!%~+p (B-p) [ k is implicit argument.]

(L())         [Start main loop running with k initially zero.]

,#,'

'(1111)       [These really are the first 4 bits of Omega!]
\end{verbatim}
}\chap{sets0.lisp}{\Size\begin{verbatim}
[[[
 Test basic (finite) set functions.
]]]

[Set membership predicate; is e in set s?]
& (Mes) /.s0 /=e+s1 (Me-s)
(Mx'(12345xabcdef))
(Mq'(12345xabcdef))
[Eliminate duplicate elements from set s]
& (Ds) /.s() /(M+s-s) (D-s) *+s(D-s)
(D'(1234512345abcdef))
(D(D'(1234512345abcdef)))
[Set union]
& (Uxy) /.xy /(M+xy) (U-xy) *+x(U-xy)
(U'(12345abcdef)'(abcdefUVWXYZ))
[Set intersection]
& (Ixy) /.x() /(M+xy) *+x(I-xy) (I-xy)
(I'(12345abcdef)'(abcdefUVWXYZ))
[Subtract set y from set x]
& (Sxy) /.x() /(M+xy) (S-xy) *+x(S-xy)
(S'(12345abcdef)'(abcdefUVWXYZ))
[Identity function that outputs a set of elements]
& (Os) /.s() *,+s(O-s)
(O'(12345abcdef))
[Combine two bit strings by interleaving them]
& (Cxy) /.xy /.yx *+x*+y(C-x-y)
(C'(0000000000)'(11111111111111111111))
\end{verbatim}
}\chap{sets1.lisp}{\Size\begin{verbatim}
[[[
 Show that
    H(X set-union Y) <= H(X) + H(Y) + c
 and that
    H(X set-intersection Y) <= H(X) + H(Y) + c
 and determine both c's.
 Here X and Y are INFINITE sets.
]]]

[Combine two bit strings by interleaving them]
& (Cxy) /.xy /.yx *+x*+y(C-x-y)

[[[++]]]?0'!%

^,#,'

[Package of set functions from sets0.lisp]
: (Mes) /.s0 /=e+s1 (Me-s)
: (Ds) /.s() /(M+s-s) (D-s) *+s(D-s)
: (Uxy) /.xy /(M+xy) (U-xy) *+x(U-xy)
: (Ixy) /.x() /(M+xy) *+x(I-xy) (I-xy)
: (Sxy) /.x() /(M+xy) (S-xy) *+x(S-xy)
: (Os) /.s() *~,+s(O-s) [<===cheating to get display!]
[Main Loop:
 o is set of elements output so far.
 For first set,
 t is depth limit (time), b is bits read so far.
 For second set,
 u is depth limit (time), c is bits read so far.
]
: (Lotbuc)
 : v     ~?~t'!%~b   [Run 1st computation again.]
 : w     ~?~u'!%~c   [Run 2nd computation again.]
 : x     (U-v-w)     [Form UNION of sets so far]
 : y     (O(Sxo))    [Output all new elements]
 [This is an infinite loop. But to make debugging easier,
  halt if both computations have halted.]
 / /=0.+v /=0.+w 100 x [If halts, value is output so far]
 [Bump everything before branching to head of loop]
 (L x                [Value of y is discarded, x is new o]
    /="?+v *1t t     [Increment time for 1st computation]
    /="!+v  ^b*@() b [Increment tape for 1st computation]
    /="?+w *1u u     [Increment time for 2nd computation]
    /="!+w  ^c*@() c [Increment tape for 2nd computation]
 )

(L()()()()())        [Initially all 5 induction variables
                      are empty.]

,
(C                   [Combine their bits in order needed]
                     [Wrong order if programs use @ or %]
,#,'*,a*,b*,c0       [Program to enumerate 1st FINITE set]
,#,'*,b*,c*,d0       [Program to enumerate 2nd FINITE set]
)
\end{verbatim}
}\chap{sets2.lisp}{\Size\begin{verbatim}
[[[
 Show that
    H(X set-union Y) <= H(X) + H(Y) + c
 and that
    H(X set-intersection Y) <= H(X) + H(Y) + c
 and determine both c's.
 Here X and Y are INFINITE sets.
]]]

[Combine two bit strings by interleaving them]
& (Cxy) /.xy /.yx *+x*+y(C-x-y)

[[[++]]]?0'!%

^,#,'

[Package of set functions from sets0.lisp]
: (Mes) /.s0 /=e+s1 (Me-s)
: (Ds) /.s() /(M+s-s) (D-s) *+s(D-s)
: (Uxy) /.xy /(M+xy) (U-xy) *+x(U-xy)
: (Ixy) /.x() /(M+xy) *+x(I-xy) (I-xy)
: (Sxy) /.x() /(M+xy) (S-xy) *+x(S-xy)
: (Os) /.s() *~,+s(O-s) [<===cheating to get display!]
[Main Loop:
 o is set of elements output so far.
 For first set,
 t is depth limit (time), b is bits read so far.
 For second set,
 u is depth limit (time), c is bits read so far.
]
: (Lotbuc)
 : v     ~?~t'!%~b   [Run 1st computation again.]
 : w     ~?~u'!%~c   [Run 2nd computation again.]
 : x     (I-v-w)     [Form INTERSECTION of sets so far]
 : y     (O(Sxo))    [Output all new elements]
 [This is an infinite loop. But to make debugging easier,
  halt if both computations have halted.]
 / /=0.+v /=0.+w 100 x [If halts, value is output so far]
 [Bump everything before branching to head of loop]
 (L x                [Value of y is discarded, x is new o]
    /="?+v *1t t     [Increment time for 1st computation]
    /="!+v  ^b*@() b [Increment tape for 1st computation]
    /="?+w *1u u     [Increment time for 2nd computation]
    /="!+w  ^c*@() c [Increment tape for 2nd computation]
 )

(L()()()()())        [Initially all 5 induction variables
                      are empty.]

,
(C                   [Combine their bits in order needed]
                     [Wrong order if programs use @ or %]
,#,'*,a*,b*,c0       [Program to enumerate 1st FINITE set]
,#,'*,b*,c*,d0       [Program to enumerate 2nd FINITE set]
)
\end{verbatim}
}\chap{sets3.lisp}{\Size\begin{verbatim}
[[[
 Show that
    H(X set-union Y) <= H(X) + H(Y) + c
 and that
    H(X set-intersection Y) <= H(X) + H(Y) + c
 and determine both c's.
 Here X and Y are INFINITE sets.
]]]

[Combine two bit strings by interleaving them]
& (Cxy) /.xy /.yx *+x*+y(C-x-y)

[IMPORTANT: This test case never halts] [<=====!!!]

[[[++]]]?0'!%

^,#,'

[Package of set functions from sets0.lisp]
: (Mes) /.s0 /=e+s1 (Me-s)
: (Ds) /.s() /(M+s-s) (D-s) *+s(D-s)
: (Uxy) /.xy /(M+xy) (U-xy) *+x(U-xy)
: (Ixy) /.x() /(M+xy) *+x(I-xy) (I-xy)
: (Sxy) /.x() /(M+xy) (S-xy) *+x(S-xy)
: (Os) /.s() *~,+s(O-s) [<===cheating to get display!]
[Main Loop:
 o is set of elements output so far.
 For first set,
 t is depth limit (time), b is bits read so far.
 For second set,
 u is depth limit (time), c is bits read so far.
]
: (Lotbuc)
 : v     ?t'!%b      [Run 1st computation again.]
 : w     ?u'!%c      [Run 2nd computation again.]
 : x     (I-v-w)     [Form INTERSECTION of sets so far]
 : y     (O(Sxo))    [Output all new elements]
 [This is an infinite loop. But to make debugging easier,
  halt if both computations have halted.]
 / /=0.+v /=0.+w 100 x [If halts, value is output so far]
 [Bump everything before branching to head of loop]
 (L x                [Value of y is discarded, x is new o]
    /="?+v *1t t     [Increment time for 1st computation]
    /="!+v  ^b*@() b [Increment tape for 1st computation]
    /="?+w *1u u     [Increment time for 2nd computation]
    /="!+w  ^c*@() c [Increment tape for 2nd computation]
 )

(L()()()()())        [Initially all 5 induction variables
                      are empty.]

,
(C                   [Combine their bits in order needed]
                     [Wrong order if programs use @ or %]
                     [Program to enumerate 1st INFINITE set]
,#,':(Lk)(L,*1*1k)(L())
                     [Program to enumerate 2nd INFINITE set]
,#,':(Lk)(L,*1*1*1k)(L())
)
\end{verbatim}
}\chap{sets4.lisp}{\Size\begin{verbatim}
[[[
 Show that
    H(X set-union Y) <= H(X) + H(Y) + c
 and that
    H(X set-intersection Y) <= H(X) + H(Y) + c
 and determine both c's.
 Here X and Y are INFINITE sets.
]]]

[Combine two bit strings by interleaving them]
& (Cxy) /.xy /.yx *+x*+y(C-x-y)

[IMPORTANT: This test case never halts] [<=====!!!]

[[[++]]]?0'!%

^,#,'

[Package of set functions from sets0.lisp]
: (Mes) /.s0 /=e+s1 (Me-s)
: (Ds) /.s() /(M+s-s) (D-s) *+s(D-s)
: (Uxy) /.xy /(M+xy) (U-xy) *+x(U-xy)
: (Ixy) /.x() /(M+xy) *+x(I-xy) (I-xy)
: (Sxy) /.x() /(M+xy) (S-xy) *+x(S-xy)
: (Os) /.s() *~,+s(O-s) [<===cheating to get display!]
[Main Loop:
 o is set of elements output so far.
 For first set,
 t is depth limit (time), b is bits read so far.
 For second set,
 u is depth limit (time), c is bits read so far.
]
: (Lotbuc)
 : v     ?t'!%b      [Run 1st computation again.]
 : w     ?u'!%c      [Run 2nd computation again.]
 : x     (U-v-w)     [Form UNION of sets so far]
 : y     (O(Sxo))    [Output all new elements]
 [This is an infinite loop. But to make debugging easier,
  halt if both computations have halted.]
 / /=0.+v /=0.+w 100 x [If halts, value is output so far]
 [Bump everything before branching to head of loop]
 (L x                [Value of y is discarded, x is new o]
    /="?+v *1t t     [Increment time for 1st computation]
    /="!+v  ^b*@() b [Increment tape for 1st computation]
    /="?+w *1u u     [Increment time for 2nd computation]
    /="!+w  ^c*@() c [Increment tape for 2nd computation]
 )

(L()()()()())        [Initially all 5 induction variables
                      are empty.]

,
(C                   [Combine their bits in order needed]
                     [Wrong order if programs use @ or %]
                     [Program to enumerate 1st INFINITE set]
,#,':(Lk)(L,*1*1k)(L())
                     [Program to enumerate 2nd INFINITE set]
,#,':(Lk)(L,*2*2*2k)(L())
)
\end{verbatim}
}\chap{godel.lisp}{\Size\begin{verbatim}
[[[
 Show that a formal system of complexity N
 can't prove that a specific object has
 complexity > N + c, and also determine c.
 Formal system is a never halting lisp expression
 that output pairs (lisp object, lower bound
 on its complexity).  E.g., (x(1111)) means
 that x has complexity H(x) greater than 4.
]]]

[ (<xy) tells if x is less than y ]
& (<xy) /.x /.y01
            /.y0(<-x-y)
(<'(11)'(11))
(<'(11)'(111))
(<'(111)'(11))

[ Examine pairs in p to see if 2nd is greater than n ]
[ returns 0 to indicate not found, or pair if found ]
& (Epn) /.p 0 /(<n+-+p) +p (E-pn)
(E'((x(11))(y(111)))'())
(E'((x(11))(y(111)))'(1))
(E'((x(11))(y(111)))'(11))
(E'((x(11))(y(111)))'(111))
(E'((x(11))(y(111)))'(1111))

++?0'!%

^,#,'
[ (<xy) tells if x is less than y ]
: (<xy) /.x /.y01
            /.y0(<-x-y)
[ Over-write real definition for test ]
: (<xy) 1

[ Examine pairs in p to see if 2nd is greater than n ]
[ returns 0 to indicate not found, or pair if found ]
: (Epn) /.p 0 /(<n+-+p) +p (E-pn)

[Parameter in proof]
: k ~'(11111)
: k ~ ^kk
: k ~ ^kk
: k ~ ^kk
: k ~ ^kk

[Main Loop - t is depth limit (time), b is bits read so far]
: (Ltb)
 : v ~?~t'!%~b [run universal computer again]
 : s (E-v^kb) [look for pair with 2nd > 16k + # of bits read]
 /s +s       [Found it!  Output 1st and halt]
 /="!+v  (Lt^b*@()) [Read another bit from program tape]
 /="?+v  (L*1tb)    [Increase depth/time limit]
 "?     [Surprise, formal system halts, so we do too]

(L()())    [Initially, 0 depth limit and no bits read]

[
,#,','((xy)(11))
]
,#,','(x())
\end{verbatim}
}\chap{godel2.lisp}{\Size\begin{verbatim}
[[[
 Show that a formal system of complexity N
 can't prove that a specific object has
 complexity > N + c, and also determine c.
 Formal system is a never halting lisp expression
 that output pairs (lisp object, lower bound
 on its complexity).  E.g., (x(1111)) means
 that x has complexity H(x) greater than 4.
]]]

++?0'!%

^,#,'
[ (<xy) tells if x is less than y ]
: (<xy) /.x /.y01
            /.y0(<-x-y)

[ Examine pairs in p to see if 2nd is greater than n ]
[ returns 0 to indicate not found, or pair if found ]
: (Epn) /.p 0 /(<n+-+p) +p (E-pn)

[Parameter in proof]
: k ~'(11111)
: k ~ ^kk
: k ~ ^kk
: k ~ ^kk
: k ~ ^kk

[Main Loop - t is depth limit (time), b is bits read so far]
: (Ltb)
 : v ~?~t'!%~b [run universal computer again]
 : s (E-v^kb) [look for pair with 2nd > 16k + # of bits read]
 /s +s       [Found it!  Output 1st and halt]
 /="!+v  (Lt^b*@()) [Read another bit from program tape]
 /="?+v  (L*1tb)    [Increase depth/time limit]
 "?     [Surprise, formal system halts, so we do too]

(L()())    [Initially, 0 depth limit and no bits read]

[
,#,','((xy)(11))
]
,#,','(x())
\end{verbatim}
}\chap{godel3.lisp}{\Size\begin{verbatim}
[[[
 Show that a formal system of complexity N
 can't determine more than N + c bits of Omega,
 and also determine c.
 Formal system is a never halting lisp expression
 that outputs lists of the form (10X0XXXX10).
 This stands for the fractional part of Omega,
 and means that these 0,1 bits of Omega are known.
 X stands for an unknown bit.
]]]

[Count number of bits in an omega that are determined.]
& (Cw) /.w() ^ /=0+w'(1) /=1+w'(1) ()
               (C-w)
(C'(XXX))
(C'(1XX))
(C'(1X0))
(C'(110))

[Merge bits of data into unknown bits of an omega.]
& (Mw) /.w() * /=0+w0 /=1+w1 @
               (M-w)
[Test it.]
++?0 ':(Mw)/.w()*/=0+w0/=1+w1@(M-w) (M'(00X00X00X)) '(111)
++?0 ':(Mw)/.w()*/=0+w0/=1+w1@(M-w) (M'(11X11X111)) '(00)

[(<xy) tells if x is less than y.]
& (<xy) /.x /.y01
            /.y0(<-x-y)
(<'(11)'(11))
(<'(11)'(111))
(<'(111)'(11))

[
 Examine omegas in list w to see if in any one of them
 the number of bits that are determined is greater than n.
 Returns 0 to indicate not found, or what it found.
]
& (Ewn) /.w 0 /(<n(C+w)) +w (E-wn)
(E'((00)(000))'())
(E'((00)(000))'(1))
(E'((00)(000))'(11))
(E'((00)(000))'(111))
(E'((00)(000))'(1111))

++?0'!%

^,#,'
[Count number of bits in an omega that are determined.]
: (Cw) /.w() ^ /=0+w'(1) /=1+w'(1) ()
               (C-w)

[Merge bits of data into unknown bits of an omega.]
: (Mw) /.w() * /=0+w0 /=1+w1 @
               (M-w)

[(<xy) tells if x is less than y.]
: (<xy) /.x /.y01
            /.y0(<-x-y)
[Over-write real definition for test.]
: (<xy) 1

[
 Examine omegas in list w to see if in any one of them
 the number of bits that are determined is greater than n.
 Returns 0 to indicate not found, or what it found.
]
: (Ewn) /.w 0 /(<n(C+w)) +w (E-wn)

[Parameter in proof]
: k ~'(11111)
: k ~ ^kk
: k ~ ^kk
: k ~ ^kk
: k ~ ^kk

[Main Loop: t is depth limit (time), b is bits read so far.]
: (Ltb)
 : v     ~?~t'!%~b  [Run universal computer again.]
 : s     (E-v^kb)   [Look for an omega with >
                     (16k + # of bits read) bits determined.]
 /s      (Ms)       [Found it!  Merge in undetermined bits,
                     output result, and halt.]
 /="!+v  (Lt^b*@()) [Read another bit from program tape.]
 /="?+v  (L*1tb)    [Increase depth/time limit.]
         "?         [Surprise, formal system halts,
                     so we do too.]

(L()())             [Initially, 0 depth limit
                     and no bits read.]

^,#,'
,'(1X0) [Toy formal system with only one theorem.]

,'
(0) [Missing bit of omega that is needed.]
\end{verbatim}
}

\part{The Software}

{
% *.c
}\chap{lisp.c}{\Size\begin{verbatim}
/* lisp.c: self-contained high-speed LISP interpreter */

/*
   The storage required by this interpreter is 8 bytes times
   the symbolic constant SIZE, which is 8 * 10,000,000 =
   80 megabytes.  To run this interpreter in small machines,
   reduce the #define SIZE 10000000 below.

   To compile, type
      cc -O -olisp lisp.c
   To run interactively, type
      lisp
   To run with output on screen, type
      lisp <test.lisp
   To run with output in file, type
      lisp <test.lisp >test.run

   Reference:  Kernighan & Ritchie,
   The C Programming Language, Second Edition,
   Prentice-Hall, 1988.
*/

#include <stdio.h>
#include <time.h>

#define SIZE 10000000 /* numbers of nodes of tree storage */
#define LAST_ATOM 128 /* highest integer value of character */
#define nil 128 /* null pointer in tree storage */
#define question -1 /* error pointer in tree storage */
#define exclamation -2 /* error pointer in tree storage */
#define infinity 999999999 /* "infinite" depth limit */

long hd[SIZE+1], tl[SIZE+1]; /* tree storage */
long next = nil; /* list of free nodes */
long low = LAST_ATOM+1; /* first never-used node */
long vlst[LAST_ATOM+1]; /* bindings of each atom */
long tape; /* Turing machine tapes */
long display; /* display indicators */
long outputs; /* output stacks */
long q; /* for converting expressions to binary */
long col; /* column in each 50 character chunk of output
            (preceeded by 12 char prefix) */
time_t time1; /* clock at start of execution */
time_t time2; /* clock at end of execution */

long ev(long e); /* initialize and evaluate expression */
void initialize_atoms(void); /* initialize atoms */
void clean_env(void); /* clean environment */
void restore_env(void); /* restore dirty environment */
long eval(long e, long d); /* evaluate expression */
/* evaluate list of expressions */
long evalst(long e, long d);
/* bind values of arguments to formal parameters */
void bind(long vars, long args);
long at(long x); /* atomic predicate */
long jn(long x, long y); /* join head to tail */
long pop(long x); /* return tl & free node */
void fr(long x); /* free list of nodes */
long eq(long x, long y); /* equal predicate */
long cardinality(long x); /* number of elements in list */
long append(long x, long y); /* append two lists */
/* read one square of Turing machine tape */
long getbit(void);
/* read one character from Turing machine tape */
long getchr(void);
/* read expression from Turing machine tape */
long getexp(long top);
void putchr(long x); /* convert character to binary */
void putexp(long x); /* convert expression to binary */
long out(char *x, long y); /* output expression */
void out2(long x); /* really output expression */
void out3(long x); /* really really output expression */
long chr2(void); /* read character - skip blanks,
                    tabs and new line characters */
long chr(void); /* read character - skip comments */
long in(long mexp, long rparenokay); /* input s-exp */

main() /* lisp main program */
{
char name_colon[] = "X:"; /* for printing name: def pairs */

time1 = time(NULL); /* start timer */
printf("lisp.c\n\nLISP Interpreter Run\n");
initialize_atoms();

while (1) {
      long e, f, name, def;
      printf("\n");
      /* read lisp meta-expression, ) not okay */
      e = in(1,0);
      /* flush rest of input line */
      while (putchar(getchar()) != '\n');
      printf("\n");
      f = hd[e];
      name = hd[tl[e]];
      def = hd[tl[tl[e]]];
      if (f == '&') {
      /* definition */
         if (at(name)) {
         /* variable definition, e.g., & x '(abc) */
            def = out("expression",def);
            def = ev(def);
         } /* end of variable definition */
         else          {
         /* function definition, e.g., & (Fxy) *x*y() */
            long var_list = tl[name];
            name = hd[name];
            def = jn('&',jn(var_list,jn(def,nil)));
         } /* end of function definition */
         name_colon[0] = name;
         out(name_colon,def);
         /* new binding replaces old */
         vlst[name] = jn(def,nil);
         continue;
      } /* end of definition */
      /* write corresponding s-expression */
      e = out("expression",e);
      /* evaluate expression */
      e = out("value",ev(e));
   }
}

long ev(long e) /* initialize and evaluate expression */
{
 long d = infinity; /* "infinite" depth limit */
 long v;
 tape = jn(nil,nil);
 display = jn('Y',nil);
 outputs = jn(nil,nil);
 v = eval(e,d);
 if (v == question) v = '?';
 if (v == exclamation) v = '!';
 return v;
}

void initialize_atoms(void) /* initialize atoms */
{
 long i;
 for (i = 0; i <= LAST_ATOM; ++i) {
 hd[i] = tl[i] = i; /* so that hd & tl of atom = atom */
 /* initially each atom evaluates to self */
 vlst[i] = jn(i,nil);
 }
}

long jn(long x, long y) /* join two lists */
{
 long z;
 /* if y is not a list, then jn is x */
 if ( y != nil && at(y) ) return x;

 if (next == nil) {
  if (low > SIZE) {
  printf("Storage overflow!\n");
  exit(0);
  }
 next = low++;
 tl[next] = nil;
 }

 z = next;
 next = tl[next];
 hd[z] = x;
 tl[z] = y;

 return z;
}

long pop(long x) /* return tl & free node */
{
 long y;
 y = tl[x];
 tl[x] = next;
 next = x;
 return y;
}

void fr(long x) /* free list of nodes */
{
 while (x != nil) x = pop(x);
}

long at(long x) /* atom predicate */
{
 return ( x <= LAST_ATOM );
}

long eq(long x, long y) /* equal predicate */
{
 if (x == y) return 1;
 if (at(x)) return 0;
 if (at(y)) return 0;
 if (eq(hd[x],hd[y])) return eq(tl[x],tl[y]);
 return 0;
}

long eval(long e, long d) /* evaluate expression */
{
/*
 e is expression to be evaluated
 d is permitted depth - integer, not pointer to tree storage
*/
 long f, v, args, x, y, z, vars, body;

 /* find current binding of atomic expression */
 if (at(e)) return hd[vlst[e]];

 f = eval(hd[e],d); /* evaluate function */
 e = tl[e]; /* remove function from list of arguments */
 if (f < 0) return f; /* function = error value? */

 if (f == '\'') return hd[e]; /* quote */

 if (f == '/') { /* if then else */
 v = eval(hd[e],d);
 e = tl[e];
 if (v < 0) return v; /* error? */
 if (v == '0') e = tl[e];
 return eval(hd[e],d);
 }

 args = evalst(e,d); /* evaluate list of arguments */
 if (args < 0) return args; /* error? */

 x = hd[args]; /* pick up first argument */
 y = hd[tl[args]]; /* pick up second argument */
 z = hd[tl[tl[args]]]; /* pick up third argument */

 switch (f) {
 case '@': {fr(args); return getbit();}
 case '%': {fr(args); return getexp('Y');}
 case '#': {fr(args);
           v = q = jn(nil,nil); putexp(x); return pop(v);}
 case '+': {fr(args); return hd[x];}
 case '-': {fr(args); return tl[x];}
 case '.': {fr(args); return (at(x) ? '1' : '0');}
 case ',': {fr(args); hd[outputs] = jn(x,hd[outputs]);
           return (hd[display] == 'Y' ? out("display",x): x);}
 case '~': {fr(args); return out("display",x);}
 case '=': {fr(args); return (eq(x,y) ? '1' : '0');}
 case '*': {fr(args); return jn(x,y);}
 case '^': {fr(args);
           return append((at(x)?nil:x),(at(y)?nil:y));}
 }

 if (d == 0) {fr(args); return question;} /* depth exceeded
                                             -> error! */
 d--; /* decrement depth */

 if (f == '!') {
 fr(args);
 clean_env(); /* clean environment */
 v = eval(x,d);
 restore_env(); /* restore unclean environment */
 return v;
 }

 if (f == '?') {
 fr(args);
 x = cardinality(x); /* convert s-exp into number */
 clean_env();
 tape = jn(z,tape);
 display = jn('N',display);
 outputs = jn(nil,outputs);
 v = eval(y,(d <= x ? d : x));
 restore_env();
 z = hd[outputs];
 tape = pop(tape);
 display = pop(display);
 outputs = pop(outputs);
 if (v == question) return (d <= x ? question : jn('?',z));
 if (v == exclamation) return jn('!',z);
 return jn(jn(v,nil),z);
 }

 f = tl[f];
 vars = hd[f];
 f = tl[f];
 body = hd[f];

 bind(vars,args);
 fr(args);

 v = eval(body,d);

 /* unbind */
 while (!at(vars)) {
 if (at(hd[vars]))
 vlst[hd[vars]] = pop(vlst[hd[vars]]);
 vars = tl[vars];
 }

 return v;
}

void clean_env(void) /* clean environment */
{
 long i;
 for (i = 0; i <= LAST_ATOM; ++i)
 vlst[i] = jn(i,vlst[i]); /* clean environment */
}

void restore_env(void) /* restore unclean environment */
{
 long i;
 for (i = 0; i <= LAST_ATOM; ++i)
 vlst[i] = pop(vlst[i]); /* restore unclean environment */
}

long cardinality(long x) /* number of elements in list */
{
 if (at(x)) return (x == nil ? 0 : infinity);
 return 1+cardinality(tl[x]);
}

/* bind values of arguments to formal parameters */
void bind(long vars, long args)
{
 if (at(vars)) return;
 bind(tl[vars],tl[args]);
 if (at(hd[vars]))
 vlst[hd[vars]] = jn(hd[args],vlst[hd[vars]]);
}

long evalst(long e, long d) /* evaluate list of expressions */
{
 long x, y;
 if (at(e)) return nil;
 x = eval(hd[e],d);
 if (x < 0) return x; /* error? */
 y = evalst(tl[e],d);
 if (y < 0) return y; /* error? */
 return jn(x,y);
}

long append(long x, long y) /* append two lists */
{
 if (at(x)) return y;
 return jn(hd[x],append(tl[x],y));
}

/* read one square of Turing machine tape */
long getbit(void)
{
 long x;
 if (at(hd[tape])) return exclamation; /* tape finished ! */
 x = hd[hd[tape]];
 hd[tape] = tl[hd[tape]];
 return (x == '0' ? '0' : '1');
}

/* read one character from Turing machine tape */
long getchr(void)
{
 long c, b, i;
 c = 0;
 for (i = 0; i < 7; ++i) {
 b = getbit();
 if (b < 0) return b; /* error? */
 c = c + c + b - '0';
 }
 /* nonprintable ASCII -> ? */
 return (c > 31 && c < 127 ? c : '?');
}

/* read expression from Turing machine tape */
long getexp(long top)
{
 long c = getchr(), first, last, next;
 if (c < 0) return c; /* error? */
 if (top == 'Y' && c == ')') return nil; /* top level only */
 if (c != '(') return c;
 /* list */
 first = last = jn(nil,nil);
 while ((next = getexp('N')) != ')') {
 if ( next < 0 ) return next; /* error? */
 last = tl[last] = jn(next,nil);
 }
 return pop(first);
}

void putchr(long x) /* convert character to binary */
{
 q = tl[q] = jn(( x &  64 ? '1' : '0' ), nil);
 q = tl[q] = jn(( x &  32 ? '1' : '0' ), nil);
 q = tl[q] = jn(( x &  16 ? '1' : '0' ), nil);
 q = tl[q] = jn(( x &   8 ? '1' : '0' ), nil);
 q = tl[q] = jn(( x &   4 ? '1' : '0' ), nil);
 q = tl[q] = jn(( x &   2 ? '1' : '0' ), nil);
 q = tl[q] = jn(( x &   1 ? '1' : '0' ), nil);
}

void putexp(long x) /* convert expression to binary */
{
 if ( at(x) && x != nil ) {putchr(x); return;}
 putchr('(');

 while (!at(x)) {
 putexp(hd[x]);
 x = tl[x];
 }

 putchr(')');
}

long out(char *x, long y) /* output expression */
{
   printf("%-12s",x);
   col = 0; /* so can insert \n and 12 blanks
               every 50 characters of output */
   out2(y);
   printf("\n");
   return y;
}

void out2(long x) /* really output expression */
{
   if ( at(x) && x != nil ) {out3(x); return;}
   out3('(');
   while (!at(x)) {
      out2(hd[x]);
      x = tl[x];
      }
   out3(')');
}

void out3(long x) /* really really output expression */
{
   if (col++ == 50) {printf("\n%-12s"," ");  col = 1;}
   putchar(x);
}

long chr2(void) /* read character - skip blanks,
                   tabs and new line characters */
{
   long c;
   do {
      c = getchar();
      if (c == EOF) {
         time2 = time(NULL);
         printf(
         "End of LISP Run\n\nElapsed time is %.0f seconds.\n",
         difftime(time2,time1)
         );
         exit(0); /* terminate execution */
         }
      putchar(c);
   }
/* keep only non-blank printable ASCII codes */
   while (c >= 127 || c <= 32) ;
   return c;
}

long chr(void) /* read character - skip comments */
{
   long c;
   while (1) {
      c = chr2();
      if (c != '[') return c;
      while (chr() != ']') ; /* comments may be nested */
   }
}

long in(long mexp, long rparenokay) /* input m-exp */
{
   long c = chr();
   if (c == ')') if (rparenokay) return ')'; else return nil;
   if (c == '(') { /* explicit list */
      long first, last, next;
      first = last = jn(nil,nil);
      while ((next = in(mexp,1)) != ')')
      last = tl[last] = jn(next,nil);
      return pop(first);
      }
   if (!mexp) return c; /* atom */
   if (c == '"') return in(0,0); /* s-exp */
   if (c == ':') { /* expand "let" */
      long name, def, body;
      name = in(1,0);
      def  = in(1,0);
      body = in(1,0);
      if (!at(name)) {
         long var_list;
         var_list = tl[name];
         name = hd[name];
         def =
         jn('\'',jn(jn('&',jn(var_list,jn(def,nil))),nil));
         }
      return
      jn(
       jn('\'',jn(jn('&',jn(jn(name,nil),jn(body,nil))),nil)),
       jn(def,nil)
      );
      }
   switch (c) {
      case '@': case '%':
         return jn(c,nil);
      case '+': case '-': case '.': case '\'':
      case ',': case '!': case '#': case '~':
         return jn(c,jn(in(1,0),nil));
      case '*': case '=': case '&': case '^':
         return jn(c,jn(in(1,0),jn(in(1,0),nil)));
      case '/': case ':': case '?':
         return jn(c,jn(in(1,0),jn(in(1,0),jn(in(1,0),nil))));
      default:
         return c;
      }
}
\end{verbatim}
}\end{document}